\documentclass[conference]{IEEEtran}

\usepackage{amsmath,amssymb} 
\usepackage{color}           
\usepackage{graphicx}
\usepackage{caption}
\usepackage{subcaption}
\usepackage{epstopdf}        
\usepackage{url}
\usepackage[ruled,linesnumbered]{algorithm2e}
\usepackage{enumerate}
\usepackage{listings,xcolor}
\usepackage{footnote}
\IEEEoverridecommandlockouts
\renewcommand\textfloatsep{-5pt}
\renewcommand\floatsep{0pt}
\makesavenoteenv{table}
\makesavenoteenv{tabular}

\title{Models and Framework for Adversarial Attacks on Complex Adaptive Systems 
}

\author{
		\IEEEauthorblockN
		{
			Vahid Behzadan\IEEEauthorrefmark{1} and  %
			Arslan Munir\IEEEauthorrefmark{2}
		}
		
		\IEEEauthorblockA
		{
			Email:                                 %
			\IEEEauthorrefmark{1}behzadan@ksu.edu and
			\IEEEauthorrefmark{2}amunir@ksu.edu
		}
  }

\begin{document}
\maketitle

\begin{abstract}
We introduce the paradigm of adversarial attacks that target the dynamics of Complex Adaptive Systems (CAS). To facilitate the analysis of such attacks, we present multiple approaches to the modeling of CAS as dynamical, data-driven, and game-theoretic systems, and develop quantitative definitions of attack, vulnerability, and resilience in the context of CAS security. Furthermore, we propose a comprehensive set of schemes for classification of attacks and attack surfaces in CAS, complemented with examples of practical attacks. Building on this foundation, we propose a framework based on reinforcement learning for simulation and analysis of attacks on CAS, and demonstrate its performance through three real-world case studies of targeting power grids, destabilization of terrorist organizations, and manipulation of machine learning agents. We also discuss potential mitigation techniques, and remark on future research directions in analysis and design of secure complex adaptive systems.
\end{abstract}


\section{INTRODUCTION}

From brains and immune systems, to societies and ecosystems, many natural phenomena are categorized as Complex Adaptive Systems (CAS). Such systems are characterized by the complex behaviors that are the emergent results of nonlinear interactions between a large number of components at different levels of system's organization \cite{mitchell2009complexity}. CAS are generally decentralized and governed by adaptive dynamics that enable their intrinsic adaptation and evolution in changing environments \cite{mesjasz2015complex}. Over the past decades, the multidisciplinary framework of CAS has been extensively applied to study natural mechanisms of emergent behavior in various domains, ranging from anatomical systems and biological behavior \cite{chiel1997brain} to social and economical systems \cite{folke2005adaptive}. 

Furthermore, the decentralized and adaptive operation of CAS has inspired numerous engineering solutions for distributed system architectures, such as smart power grids \cite{amin2005toward}, autonomous navigation \cite{ordoukhanian2016resilient}, and the Internet of Things (IoT) \cite{mittal2017simulation}. Equipment of such distributed systems with CAS-inspired mechanisms is a promising approach to the challenging task of control and management of the increasingly complex and heterogeneous systems \cite{athreya2013network}. In particular, the \emph{Self-organization} aspect of CAS enables the emergence of order and pattern from uncoordinated actions of autonomous agents in multi-agent distributed settings \cite{behzadan2017game}. In self-organizing systems, individual agents are capable of adapting to changes in the environment via autonomic tuning of their configurable parameters to enhance individual as well as global operations of dynamic distributed systems. 


The growing interest in adoption of CAS architectures in mission-critical applications intensifies the need for investigating the security aspects of such systems. While the distribution of responsibilities and capabilities among multiple agents in CAS seemingly relieves the threats posed by single points of failure, the complexity of dynamics in such systems gives rise to unique challenges in quantifying and ensuring their resilience and robustness in hostile environments and adversarial conditions. While the body of work on CAS presents many contributions towards analysis of resilience against random and natural perturbations, current state of the art leaves major gaps in understanding and enhancement of resilience against targeted attacks and adversarial actions.

This paper aims to develop a comprehensive foundation for analysis and enhancement of resilience in natural and engineered CAS against adversarial actions. To this end, we study and formalize the threats posed by attacks targeting the adaptive dynamics of such systems. Accordingly, the main contributions of this paper are as follows:
\begin{enumerate}
	\item We introduce three approaches to the modeling of CAS, namely: dynamical systems model, Dynamic Data-Driven Application Systems (DDDAS) model, and game theoretic model of strategic network formation.
	\item We propose quantitative definitions of attack, vulnerability, and resilience in the context of CAS security. 
	\item We develop a comprehensive set of schemes for classification of attack surfaces in CAS, and discuss generic instances of active and passive adversarial actions targeting these surfaces.
	\item We propose a framework based on reinforcement learning for simulation and analysis of attacks on CAS.
	\item We demonstrate the practical application of our proposed framework in three practical case studies: induction of cascade failures in power grids, destabilization of terrorist organizations, and manipulation of deep reinforcement learning agents.
	\item We present a discussion on potential defensive and mitigation techniques.
\end{enumerate}

The remainder of this paper is organized as follows: Section \ref{Sec:Background} provides an overview of CAS and the relevant background. Section \ref{Sec:Model} presents models for analysis of CAS. Section \ref{Sec:Vuln} details our proposed definitions of attack, vulnerability, and resilience. Section \ref{Sec:Surf} presents classifications of vulnerabilities and attack surfaces in CAS, followed by the proposal of a framework for simulation of adversarial actions and analysis of their impact on CAS in Section \ref{Sec:Framework}. Section \ref{Sec:Case} demonstrates the application of this framework in three practical case studies. Section \ref{Sec:Defense} discusses potential approaches towards mitigation of such attacks, and Section \ref{Sec:Conclusion} concludes the paper with remarks on future research directions.

\section{Background}\label{Sec:Background}
In this section, we briefly introduce the paradigm of complex systems and their adaptivity to provide the reader with an overview of fundamental concepts and notions required for the remainder of this paper. It must be noted that this background is by no means comprehensive, and the interested reader may refer to sources such as \cite{miller2009complex} and \cite{strogatz2014nonlinear} for in-depth introductions to CAS.

\subsection{Complex Adaptive Systems}
Complexity, as a quantifiable measure, is yet to obtain a unified and consistent definition. From the multitude of definitions that have emerged from the field of complexity science \cite{lloyd2001measures}, we abide by the definition presented by Mitchell \cite{mitchell2009complexity}: ``A complex adaptive system is a system in which large networks of components with simple rules of operation and no central control give rise to complex collective behavior, sophisticated information processing, and adaptation. Such systems exhibit nontrivial emergent and self-organizing behaviors.'' 

Accordingly, the most general characteristics of CAS are identified as \cite{mesjasz2015complex}:
\begin{itemize}
	\item Large numbers of constituent elements and interactions
	\item Non-decomposability, i.e., components cannot be separately studied due to interactions
	\item Nonlinearity of dynamics and behavior
	\item Various forms of hierarchical structure
	\item Emergent behavior
	\item Self-organization
	\item Co-evolution with other complex entities or the environment.	
\end{itemize}  

The concepts of emergence and self-organization are of particular significance in the scope of our work. \emph{Emergence} in CAS refers to the occurrence of properties and behavior in a system that are not present in the constituent components, i.e., global properties resulting from local interactions are emergent \cite{fernandez2014information}. Similarly, \emph{Self-Organization} is the emergence of global coherence out of local interactions \cite{marinescu2016complex}. Natural instances of self-organization include the swarming formation of birds in flight, and the emergence of cognitive abilities from interactions of neurons in the brain. 

\subsection{Vulnerability and Resilience of CAS}
The resilience of complex systems has been the subject of active research in diverse disciplines, ranging from ecology \cite{walker2004resilience} and epidemiology \cite{horwitz2005parasites} to power distribution systems \cite{pagani2013power} and counter-terrorism \cite{CT1}. Yet, the bulk of available literature on this topic emphasize on resilience of CAS to naturally occurring and random perturbations. Amid the spectrum of definitions considered in such works \cite{hosseini2016review}, one of the most general definitions of resilience is: ``The ability of a system to endure failure and recover from mishaps by restoring its capacities'' \cite{hollnagel2007resilience}. While this definition captures the objectives of system-level studies, it fails to satisfy the requirements of security analyses. While recovery from failure may demonstrate the long-term sustainability of system's operations, the security consequences of short-term failures may be catastrophic. For instance, exposure of confidential information in a cloud computing platform, however technically recoverable, may incur severe damages to the users and operators of the platform. Therefore, there is a need for security-oriented alternatives of this definition.

Similarly, the concept of vulnerability in CAS is defined either too loosely, or too context-dependent. For instance, \cite{pedroni2016advanced} defines vulnerability as the system's inability to resist stresses, which may be exploited by threats and hazards. On the other hand, \cite{moore2016quantifying} provides a network-oriented definition as links or nodes whose removal adversely impact the functions of a complex network. In the context of disaster mitigation, \cite{dalziell2004resilience} defines vulnerability as ``the human product of any physical exposure to a distater that results in some degree of loss.'' It is evident that a generic and quantitative definition of vulnerability is needed to form the basis of a quantitative framework for security analysis of CAS.

In Section \ref{Sec:Vuln}, we utilize the dynamical model of CAS to develop such definitions of resilience and vulnerability for analysis of security in such systems.

\section{Models of CAS} \label{Sec:Model}
In this section, we present three approaches to modeling the behavior of CAS. First, we introduce the dynamical system model and the relevant terminology, which will form the basis of defining resilience, vulnerability, and attack in Section \ref{Sec:Vuln}. This approach is complemented by the Dynamic Data-Driven Application System (DDDAS) abstraction, as well as a game-theoretic model of network formation. Having multiple approaches enables various levels of abstraction for high-dimensional CAS, thereby providing multiple perspectives for capturing the structure and dynamics of such systems. These approaches are detailed below. 

\subsection{Dynamical Model} \label{Sec:Dyn}
CAS are dynamical systems, meaning that their states change as a function of time. In this perspective, the dynamics of CAS can be modeled as:

\begin{eqnarray} \label{eq:dyn}
\dot{x}(t) = f(x(t), \beta(t))
\end{eqnarray}

Where $\dot{x}(t)$ is the first-order derivative of $x$ with respect to $t$, $x = (x_1, x_2, ..., x_n)$ is the $n$-dimensional state of CAS, $\beta$ is the state of the environment (or alternatively, control input), and $f$ is the dynamics of the system. The set of all possible configurations of $x$ is termed the \emph{phase space} of the system, henceforth denoted by $X$. A solution $x(t)$ to the equation \ref{eq:dyn} constitutes a \emph{trajectory} in phase space. Any trajectory is uniquely defined by the initial conditions, $x(0) \equiv x_0$. Accordingly, the Time-T \emph{Flow} $\phi_T$ for initial conditions $x(0)$ is defined as $\phi_T(x(0)) = x(T)$.

In dynamical systems, an \emph{attractor} is a bounded region in phase space to which trajectories with certain initial conditions come arbitrarily close. Formally, an attractor is an invariant set $\Lambda \in X$, where trajectories of perturbations that lead to states outside of $\Lambda$ eventually return to $\Lambda$. Attractors may be isolated points, limiting cycles, or more complex objects in the phase space.

A \emph{basin of attraction} $\Omega(\Lambda)$ is the set of all states which fall on trajectories that lead to attractor $\Lambda$. Formally,

\begin{eqnarray}
\Omega(\Lambda) = \{x\in X: \lim_{t\rightarrow \infty}\phi_t(x)\in \Lambda\}
\end{eqnarray}

Accordingly, the \emph{basin boundary} $\partial \Omega$ of a CAS is defined as the set of states that are not in any basin. Formally:
\begin{eqnarray}
\partial \Omega = X - \bigcup_{i} \Omega(\Lambda^i)
\end{eqnarray} 

Even though the dynamical model provides a fundamental mathematical perspective on the behavior of CAS, the abstraction and computational aspects of this model become severely restricted in high-dimensional systems. Therefore, alternative models are often used to simplify the dynamical representation and abstraction of CAS. 

\subsection{DDDAS Model}\label{Sec:DDDAS}
The decentralized adaptive behavior of CAS implies the existence of a feedback control loop in the constituent components. Accordingly, each component of CAS monitors the changes in its environment, analyzes the observations and its internal state with respect to local rules and objectives, and adjusts its operating parameters accordingly. This process can be accurately captured within the framework of Dynamic Data-Driven Application System (DDDAS).  A DDDAS is a symbiotic feedback control system, which can dynamically analyze the state of system and its environment to control and determine when, where, and how it is best to gather additional data, and in reverse, can dynamically steer the applications based on the obtained measurements \cite{fujimoto2016dynamic}. The operational cycle of an agent in a generic distributed DDDAS comprises of 4 components:

\begin{itemize}
	\item \emph{Sensing}: Observing the state of agent's environment and retrieving relevant information that may be disseminated by other agents
	\item \emph{Information Sharing}: Communicating agent's current state and observations with other agents
	\item \emph{Data Fusion and Analytics}: Integration and processing of observed and retrieved information
	\item \emph{Self-Configuration}: Configuration of agent's functional parameters according to processed information
\end{itemize}

\begin{figure}[h]
	\includegraphics[width=\linewidth]{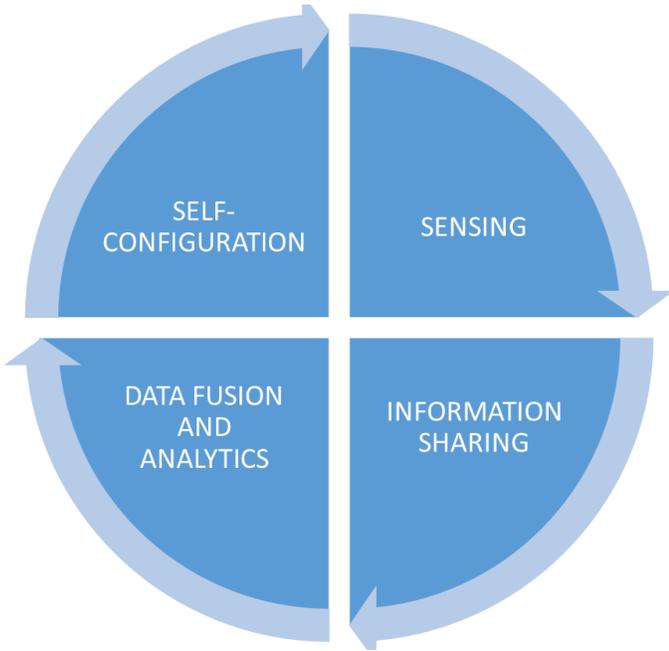}
	\caption{Operational cycle of distributed DDDAS}
	\label{FigDDDAS}
\end{figure}

Figure \ref{FigDDDAS} illustrates the anatomy of a DDDAS cycle.

Since the inception of DDDAS, this framework has spawned numerous applications such as environment analysis (e.g., weather \cite{patrikalakis2004towards}); robotic systems (e.g., coordination and swarming of unmanned aerial vehicles (UAVs) \cite{mccune2013swarm} and unmanned ground vehicles (UGVs) \cite{khaleghi2013dddams}); image processing (e.g., target tracking \cite{blasch2015dynamic}), and embedded computing (e.g., hardware/software designs \cite{sudusinghe2014model}). Furthermore, recent literature illustrates the application of this framework to the analysis of generic complex systems \cite{kutz2016dynamic}. 

\subsection{Network Formation Game Model}

CAS are networks comprised of a large number of various agents, each with unique requirements and capabilities, leading to heterogeneity in various aspects of the systems. Each individual agent in this network aims to optimize its local objectives, such as energy consumption, computational performance, reliability, and resilience, through interactions with other agents in the network. The actions and interactions of these agents give rise to emergent patterns at macro-scale, which drive the system-level behavior of self-organizing networks. 

To enable the analysis of emergent behaviors, stability, and resilience of CAS, one approach is to model the dynamics of interactions as strategic network formation games \cite{vannetelbosch2015network} that provides a framework for analysis of self-organizing dynamics for generic designs and applications. In such games, every agent desires to establish the optimal set of links to other agents which maximizes the agent's reward or utility. Depending on system specifications, a link in this setting may represent inter-node communications, routing hop, information sharing, computation and communication resource sharing, synchronized actuation, proximity, trust, or any other quantifiable relationship. Accordingly, the dynamics of interactions can be captured by a network formation game $\Gamma(N, U, \mathbb{A}, (G, F))$ with complete or incomplete information, where $N$ is the set of all agents, $\mathbb{A}$ is the set of all actions available to players, $U=\{U_1, U_2,...,U_N\}$ is the set of each agent's payoff function, $G(N, E)$ is the graph of $N$ vertex with the directional or undirectional weighted edge-set $E$ representing the network topology, and $F = \{F_1, F_2, ..., F_n\}$ is the set of attribute vectors representing the exogenous features and characteristics of each individual. The tuple $(G, F)$ is the information available to all players on the game settings. Each agent $i \in N$ also bears an idiosyncratic profile $\varepsilon_i$, capturing the traits and characteristics of individuals that affect their decision in link establishment, but are not known to other agents. Such characteristics may include experience and learning profile, priority of objectives, and level of trust. The actions of players in this game are their establishment or removal of heterogeneous links to other players. Let $G_{ij}$ be the $ij$-th component of the adjacency matrix of the network. The action of player $i$ is denoted by $G_i = ((G_{i,1}, E_{i,1}), (G_{i,2}, E_{i,2}), ..., (G_{i,\left|N\right|}, E_{i,\left|N\right|}))\in \mathbb{A}\subset m_i\times [w \in W_i \subseteq R]^{\left|N\right|}$ where $m_i$ is the number of link types available to player $i$ and $W_i$ is the set of possible link weights for $i$. 

\begin{figure}[h]
	\includegraphics[width=\linewidth]{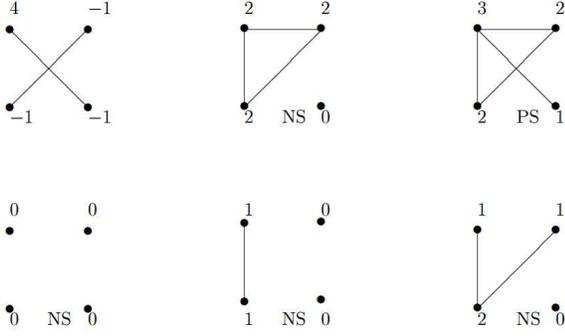}
	\caption{Disjoint criteria for stability in network formation games. Numerical values represent the payoff of each node, NS: Nash-Stable, PS: Pairwise-Stable}
	\label{FigNetGame}
\end{figure}

\begin{figure*}[]
	\centering
	\includegraphics[width=\textwidth]{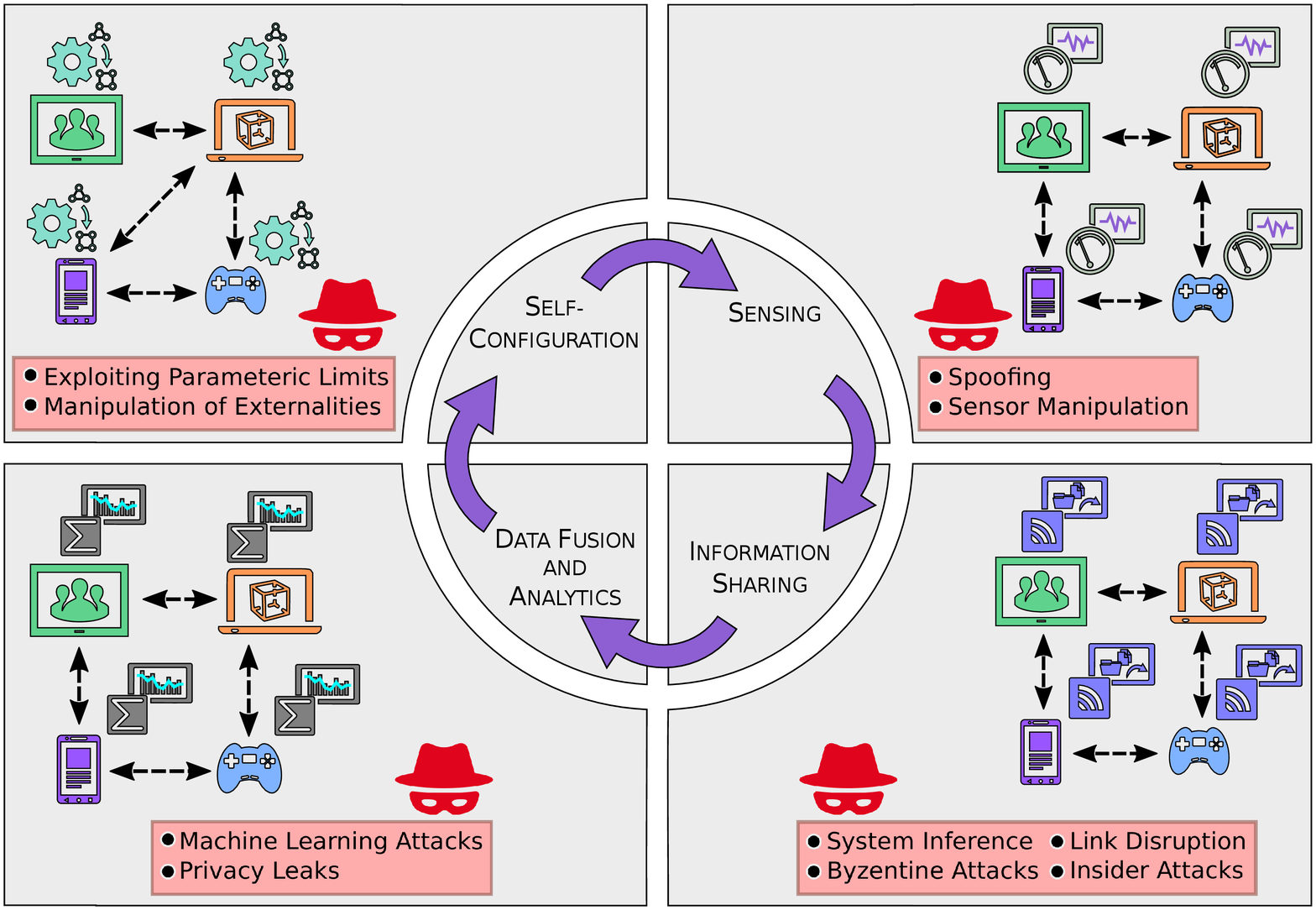}
	\caption{Instances of potential attacks on data-driven self-organizing systems.}
	\label{FigAttacks}
\end{figure*}

Various types of equilibria and stability can be defined for such games, including Nash Stability (NS) and Pairwise Stability (PS) \cite{bloch2006definitions}. As illustrated by the example in Figure \ref{FigNetGame}, these different criteria for stability do not necessarily overlap and need to be chosen according to the problem at hand. By choosing the relevant criteria for stability and defining suitable payoff functions $U_{i\in N}$ to account for relevant costs and incentives of game states and trajectories, this model allows for analysis of generic parametric bounds and relations in establishment of emerging topologies, behaviors, and dynamical stability within the game abstraction. Furthermore, this game theoretic modeling of self-organizing behavior provides the formal analysis of behaviors and interactions by considering the adversary as another player in the game. Also, network formation games can enjoy the benefits of many strong analytical toolsets such as graph theory, category theory, network science, and cooperative optimal control.

\section{Threat Model} \label{Sec:Vuln}
The adaptive dynamics of CAS gives rise to a variety of vulnerabilities and attack surfaces. By definition, the macro-scale behavior of such systems is the emergent result of micro-scale actions of local or individual elements. Therefore, adversarial perturbations of micro-scale structure and dynamics may result in amplification of perturbations and manipulation of the macro-scale behavior. 

To ensure a consistent and comprehensive study of such attacks, we first develop suitable definitions of attack, vulnerability, and resilience in CAS. We differentiate between two types of attacks, namely passive and active attacks. \emph{Passive attacks} aim at exposure of structural and dynamical properties of the targeted CAS, and do not require exertion of additional input to the system. Instances of such attacks are traffic analysis \cite{behzadan2016real} and inference of dynamics \cite{CT1}. On the other hand, \emph{Active attacks} involve the implementation of adversarial actions to achieve an adversarial objective. Building on the dynamical model of Section \ref{Sec:Dyn}, we define \emph{adversarial action} as the intentional manipulation of either the state or dynamics of CAS, such that the resulting state-space trajectory passes through states, which may include states outside of desired basins of attraction, states within undesired submanifold of the phase space (e.g., undesired basins of attraction), or ill-defined states within a modified phase space. Accordingly, the modes of adversarial actions can be categorized as those perturbing the state configuration of CAS, and those that manipulate the dynamics of CAS, formalized as follows:

\subsubsection{State Manipulation}
Let $\gamma(x_t)$ be the perturbation to state $x_{t} \in X$, i.e., the perturbed state is obtained via $x^p_{t} = x_{t} + \gamma(x_t)$. The problem of adversarial state manipulation is to devise the function $\gamma (x_t)$ such that at an arbitrary time $T$:
\begin{eqnarray}
x_T = \int_{{t_0}}^{T} f(x_t + \gamma(x_t), \beta_t)dt \in X^{*}
\end{eqnarray}

Where $t_0$ is the initial time, and $X^{*} \in X^{'}$ is the set of states within the space of undesired states $X^{'}$ which conform to adversarial objectives. It is noteworthy that a sustainable impact is imposed when the adversary aims for driving the target into $X^{*}$'s basins of attraction.

Alternatively, if the objective is to reach specific trajectories $\mu(t)$ in the space of undesired trajectories $M$ rather than particular states, the problem can be rearranged as devising $\gamma (x_t)$ s.t. some measure of distance between the original and desired trajectory becomes smaller than an arbitrary error threshold $\epsilon$, i.e.,
\begin{eqnarray}
\lVert \dot{x}(t) - \dot{\mu}(t)\rVert = \lVert F(x_t+\gamma(x_t), \beta_t) - \dot{\mu}(t) \rVert < \epsilon\\
\nonumber
\end{eqnarray}

\subsubsection{Dynamics Manipulation}
Let $\lambda(x_t, \beta_t)$ be the perturbation to the environment (alternatively, it can be viewed as control input). The problem of adversarial dynamics manipulation is to devise a suitable control perturbation $\lambda(x_t, \beta_t)$, such that at an arbitrary time $T$:

\begin{eqnarray}
x_T = \int_{{t_0}}^{T} f(x_t, \beta_t + \lambda(x_t, \beta_t))dt \in X^{*}
\end{eqnarray}

It must be noted that $X^{*}$ is not necessarily a subset of $X$, as the phase space may shift due to perturbations. Alternatively, the problem of reaching specific trajectories can be formulated similarly to the case of state manipulation, with the following optimization objective:
\begin{eqnarray}
\lVert \dot{x}(t) - \dot{\mu}(t)\rVert = \lVert F(x_t, \beta_t+\lambda(x_t)) - \dot{\mu}(t) \rVert < \epsilon\\
\nonumber
\end{eqnarray}

With the concept of attack formalized, we can construct suitable measures of vulnerability and resilience on the same grounds. We adopt a well-established fact from the realm of cyber-security that no system can be completely secure against all possible attacks. Hence, the objective of securing a system becomes deterrence of attacks in an economic sense, namely making successful attacks as costly as possible \cite{moore2010economics}. Accordingly, we define the \emph{vulnerability} of an element (state, trajectory, or dynamics) in a CAS to a specific adversarial action, as the inverse of the minimum amount of cost incurred to the adversary to impose the maximum achievable cost to the targeted CAS, via implementing the adversarial action on the designated element. This definition assumes that adversarial cost $C_{adv} \geq 1$, and hence the value of vulnerability is in the range $\lbrack 0, 1 \rbrack$, whose unit is determined by the dimensions of adversarial cost $C_{adv}$. In a similar manner, we define the \emph{resilience} of a CAS against a certain attack as the minimum cost imposed on the adversary to successfully implement that adversarial action and force the CAS into an undesired state or trajectory. The selection of adversarial and CAS cost metrics is highly dependent on the context of analysis. One simple instance of choices for adversarial cost can be the minimum number of perturbations required for a successful attack. A similar choice for the CAS cost is the loss of connectivity in the network model of its interactions. 

\section{Classification of Attack Surfaces}\label{Sec:Surf}
Attack surfaces are structural and dynamical components of CAS that may be targeted in active and passive attacks. In this section, we present three schemes categorizing such components, and provide attack instances for each identified component.

\subsubsection{CIA-based} The first approach concentrates on the security dimensions being attacked. The general dimensions of security are Confidentiality, Integrity, and Availability, forming the CIA triad of security \cite{baldwin2017contagion}. \emph{Confidentiality} refers to the restriction of unauthorized access to protected information. Examples of attacks on confidentiality in CAS include the inference of states, dynamics, and interaction protocols in a self-organizing swarm of UAVs. \emph{Integrity} is maintaining and assuring the accurate functioning of the system in the intended manner. An instance of corresponding attacks is manipulation of a distributed autonomous navigation system to induce collisions. \emph{Availability} is assuring the uninterrupted operation of the system. Induction of cascade failures in power distribution systems is a well-established instance of such attacks on CAS.


\subsubsection{DDDAS-based} Another approach to classification of attack surfaces is based on the distributed DDDAS model presented in Section \ref{Sec:DDDAS}. As illustrated in Figure \ref{FigAttacks}, each component of the DDDAS cycle constitutes attack surfaces that can be the subject of adversarial actions targeting one or a combination of the CIA dimensions. However, as shown in Table \ref{Attacks}, under this schemes some attacks may find overlapping roots between different component.

\subsubsection{Functionality-based} We also propose a more general functionality-based approach to classification. The building blocks of CAS are its structure and topology, dynamics of interactions, and the internal dynamics of each constituent agent. Accordingly, we further categorize the attack surfaces of CAS into those stemming from the \emph{Network Structure, Cooperation Protocolos,} or \emph{Actuation Functions}, detailed below:

\subsection{Attacking the Network Structure}\label{Net}
\begin{figure}[h]
	\includegraphics[width=\linewidth]{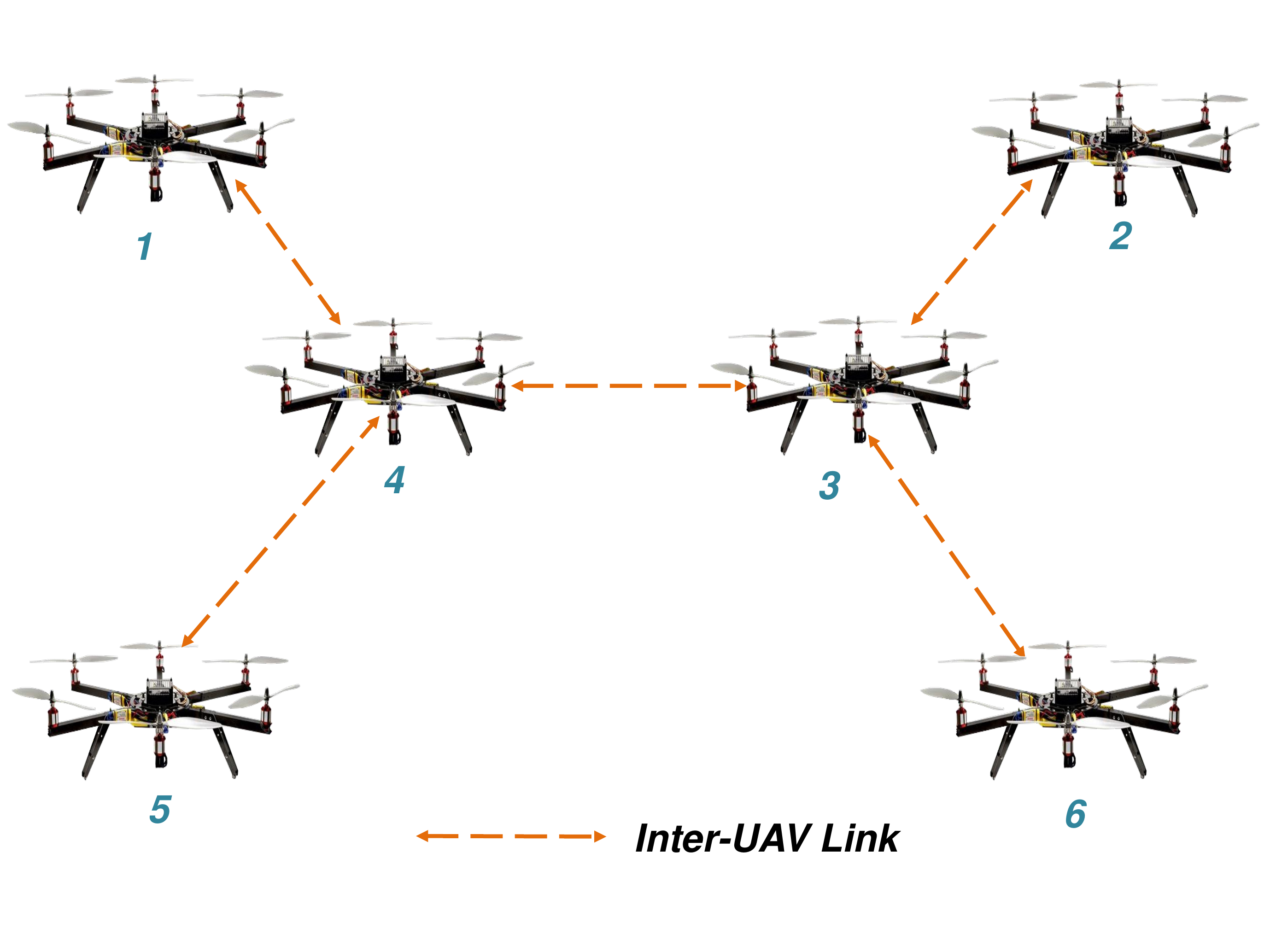}
	\caption{Example of topological vulnerability}
	\label{FigTopology}
\end{figure}

As discussed in Section \ref{Sec:Model}, CAS can be modeled as networks of interacting agents. Depending on the model's context and objective, this network may represent the communication links between agents, their interactions, dependencies, or other types of relationships. As is the case with distributed networked systems, such as communications (e.g., \cite{lopez2017vulnerability}) and social networks (e.g., \cite{wood2017structure}), the intrinsic network structure of CAS gives rise to a number of potential vulnerabilities that can be exploited to mount passive and active attacks against the system. By means of \emph{traffic analysis} \cite{behzadan2016real} and \emph{inference} attacks \cite{CT1}, adversaries can target the confidentiality of CAS to identify the topology and dynamics of their networks. Knowledge of the network topology allows adversaries to optimize denial of service attacks by analyzing the structure of their target and determining the most critical regions \cite{behzadan2016real}. To further expand on this surface, consider the case of a self-organizing swarm of UAVs, as illustrated in Figure \ref{FigTopology}. The inter-UAV network depicted in this figure is a graph with 2 hubs (i.e., Nodes 3 and 4), through which a large portion of network flows pass. If the adversary aims a jamming attack at only these two hubs, the network becomes completely disconnected, thereby the entire operation of the system is disrupted at minimal cost to the adversary. Under certain circumstances, this type of attack may cause cascading effects that result in total system failure over time. A well-studied example of which is cascade failures in power grids \cite{guo2017critical}. 

\subsection{Attacking Cooperation Protocols}
Considering the independent and self-interested nature of agents in CAS, stabilization and efficiency of many real-world applications of such systems necessitate the implementation of rules and protocols to induce and maintain cooperative interactions between agents. For instance, formation control and navigation of UAV swarms require the sharing of positional information among UAVs, as well as their coordination of navigational parameters. Implementation of cooperation protocol creates another source of attack surfaces. Adversaries may target the confidentiality of CAS via passive sniffing of shared information through either insider and outsider attacks. This type of passive eavesdropping enables further active attacks through inference and identification of objectives and system dynamics. 

The integrity of such systems can be targeted in various ways. By spoofing legitimate agents, adversaries can inject false data into the information sharing pipeline of CAS. Also, spoofed, compromised, or malicious insider agents may falsify their resource requirements, or even pose as several agents to gain unfair access to shared resources. In the domain of distributed wireless networks, this type of exploitation is known as \emph{Sybil attack} \cite{pathan2016security}. Furthermore, in systems with constrained information sharing capacities, adversarial perturbation of the environment may lead to sharing of incorrect or incomplete information. For instance, consider the case of a UAV swarm which relies on individual reporting of observed obstacles for collision avoidance. If the reporting protocol limits the number of reported obstacles to the $n$ nearest objects observed by a UAV, an adversary may spoof or generate $m >> n$ minor obstacles in the vicinity of the UAV to prevent it from informing rest of the swarm about major nearby obstacles.

Attacks on the availability aspect may also come in different forms. Spoofed, compromised, or malicious insider agents may act as information \emph{blackholes} \cite{sarma2014survey} by tactically refusing to share their information at particular times. In CAS that rely on multi-hop communications, this attack can be more damaging if the agent stops forwarding information received from other neighbors as well. Another type of attack is based on spoofed, compromised, or malicious insider agents disseminating certain information that cause termination of cooperation. In our example of UAV swarm, transmission of messages such as ``mission accomplished'', ``mission failed'', or radio silence signal in tactical scenarios, may cause the cooperative process to end. Furthermore, if the cooperation protocol is not well-designed, broadcast of certain resource constraints or environmental conditions may result in prevalence of agents' selfishness over cooperation. For instance, if the UAV swarm encounters an inevitable collision state \cite{behzadan2017cyber}, cooperation protocol may allow agents to choose independent action over cooperation. This condition may be induced through either dissemination of fake information, or adversarial manipulation of the environment.

\subsection{Attacks on Actuation Functions}
The main objectives of CAS are realized by each agent via actuation functions. In the example of UAVs, actuation functions are cyber-physical controllers of motion and communications. In general, the ultimate goal of all attacks introduced so far is indirect manipulation or disruption of actuation functions. Adversaries may also directly target the actuation of CAS through attack surfaces in actuation mechanisms and functions. Mounting attacks on confidentiality of actuation may be in the form of parameter inference. Obtaining knowledge of operating parameters through side-channel attacks enables the adversary to derive a more accurate estimation of system's state and dynamics, thereby allowing the optimization of active attacks against the system. Also, in competitive CAS, complete knowledge of an agent's operating parameters may provide other agents with an unfair advantage. For instance, consider a CAS setup to automate the sharing of information on cyber attacks among corporations \cite{vakilinia2017evolving}. In this scenario, agents aim to share the minimal amount of data required to preserve the long-term benefits of information sharing. If an adversarial agent is able to estimate the parameters used by another agent in filtering and disseminating information, it may allow the adversary to infer the undisclosed portion of agent's information. A sophisticated attack in such incomplete information systems can be the adversarial disclosure of parameters to competitors, thereby causing the system dynamics to
diverge from a beneficial equilibrium. Economic and political parallels of this phenomenon are instances of insider trading and whistleblowing (e.g., \cite{smales2017game}).

The integrity of actuation functions may be targeted via manipulation of the environment or sensory observations. In an autonomous fleet of self-organizing vehicles, calculated manipulation of the visual input to a vehicle may result in an \emph{adversarial example} \cite{papernot2016practical} for the machine learning component of the system. Adversarial examples are minimally perturbed inputs that cause misclassifications in machine learning algorithms. For instance, minor changes in a speed sign on the side of a street can result in its misclassification as a stop sign by an autonomous vehicle, causing it to stop in an unsafe location \cite{behzadan2017cyber}. In some cases, even spoofed perturbations of the environment is sufficient for manipulation of actuation functions. A real-world example of such cases is the Automatic Collision Avoidance System (ACAS) utilized by many of today's commercial aircraft \cite{kastelein2014preliminary}. This system generates motion advisories according to the position and heading of other aircraft in the environment, obtained from an unencrypted, open protocol known as ADS-B \cite{strohmeier2013security}. An adversary may simply fake the presence and trajectory of nonexistent aircraft by spoofing, ADS-B signals, which can lead to ACAS advisories that change the trajectory of targeted aircraft \cite{behzadan2017cyber}.

\begin{figure}[h]
	\includegraphics[width=\linewidth]{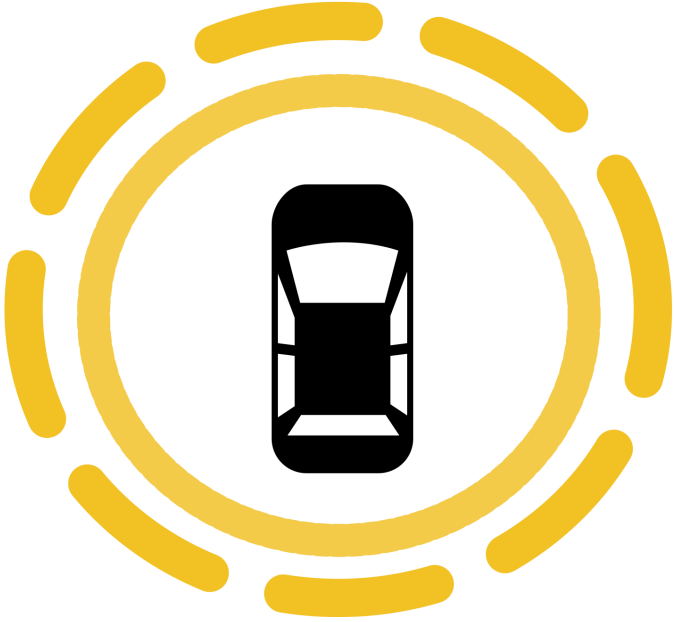}
	\caption{Denial of service via manipulation of the environment}
	\label{FigAuto}
\end{figure}

Similar attacks can also target the availability of actuation functions. Adversaries may manipulate the environment such that the actuation functions of CAS agents fall within undefined or terminal states. Figure \ref{FigAuto} illustrates an instance of such attacks: an autonomous vehicle that is trained to avoid crossing solid lines, will inevitably remain stationary if it finds itself encircled by such lines. In our UAV example, induction of emergency conditions through environmental or sensory manipulation can drive targeted agents into safe modes, which in many cases trigger automatic Return-to-Base (RTB) or emergency landing procedures \cite{behzadan2017cyber}.

Table \ref{Attacks} presents the classifications of the sample attacks discussed in this section.

\begin{table*}[t]
	\centering
	\fontsize{9}{9}\selectfont
	\begin{tabular}{llcccc}
		Functional Surface    & Attack Example                       & \multicolumn{1}{l}{CIA Dimension} & \multicolumn{1}{l}{DDDAS Surface} & \multicolumn{1}{l}{Attack Type} & \multicolumn{1}{l}{Attack Mode} \\ \hline \\
		Network Structure     & Traffic Analysis, Topology Inference & C                                 & IS                                & Passive                         & N/A                             \\
		& Topological Disruption               & A                                 & IS                                & Active                          & State                           \\
		& Cascade Induction                    & I, A                              & IS, SC                            & Active                          & Dynamics                        \\
		&                                      & \multicolumn{1}{l}{}              & \multicolumn{1}{l}{}              & \multicolumn{1}{l}{}            & \multicolumn{1}{l}{}            \\
		Cooperation Protocols & Sniffing                             & C                                 & IS                                & Passive                         & N/A                             \\
		& Sybil                                & I, A                              & IS, SC, S                         & Active                          & State/Dynamics                  \\
		& Information Manipulation             & I, A                              & SC, AN                            & Active                          & Dynamics                        \\
		&                                      & \multicolumn{1}{l}{}              & \multicolumn{1}{l}{}              & \multicolumn{1}{l}{}            & \multicolumn{1}{l}{}            \\
		Actuation Functions   & Parameter/Dynamics Inference         & C                                 & IS, SC                            & Passive                         & N/A                             \\
		& Competitive Intelligence             & C                                 & IS, SC, AN                        & Passive                         & N/A                             \\
		& Adversarial Examples                 & I, A                              & S, AN, SC                         & Active                          & State                           \\
		& Spoofing                             & I, A                              & S, AN, SC                         & Active                          & State / Dynamics                \\
		& Induction of Terminal States         & I, A                              & S, AN, SC                         & Active                          & State \\ \hline                         
	\end{tabular}
	\caption{Classification of sample attacks - C, I, and A stand for Confidentiality, Integrity and Availability, respectively. For DDDAS attack surfaces, S is Sensing, IS is Information Sharing, AN stands for Analytics, and SC is Self-Configuration.}
	\label{Attacks}
\end{table*}

\section{Simulation Framework} \label{Sec:Framework}
As an approach towards analysis of impact in attacking CAS dynamics, we propose a framework for simulation of adversarial actions against generic CAS. With the aim of analyzing the maximum impact of attacks, this framework is designed automatically derive the optimal sequence of adversarial actions against CAS models. Also, our framework supports the analysis of both whitebox and blackbox attacks, meaning that the adversary can be considered to have complete, partial, or no a priori knowledge of the system dynamics. Furthermore, this framework allows for arbitrary designation of adversarial goals (e.g., network disruption, actuation manipulation, etc.), and can be configured for arbitrary types of adversarial actions.

The initial step of each simulation in this framework is to obtain an estimation of dynamics in the targeted CAS from time-series observations of the system. For simulation of blackbox attacks, this can be achieved through a variety of methods developed for identification of nonlinear dynamics, such as utilization of deep neural network (e.g., \cite{ogunmolu2016nonlinear}). When partial knowledge of the system is assumed, the estimation technique can be based on a generic model of the dynamics with unknown model parameters, which may be estimated via statistical techniques. As for the simulation of whitebox attacks, this estimation can be fixed to a complete dynamical model of the system. Examples of each case are presented in Section \ref{Sec:Case}.

With the initial estimate of dynamics at hand, the next step of this framework is to create a secondary simulation of the targeted system in order to obtain the optimal attack policy $\pi^*(S)$, which maps any observed state $S$ of the estimated system to an optimal action $A_S$. This action corresponds to one the adversarial actions defined in the initial configuration of simulations, Instances of which are node removals for attacks on network structure, sensory overload for attacks on cooperation protocols, and crafting adversarial examples for manipulation of actuation functions. 

Accordingly, we propose \emph{reinforcement learning} as a promising approach to the problem of policy optimization. Reinforcement learning techniques are described by the Markov Decision Process (MDP) tuple $(S, A, P, R)$, where $S$ is the set of reachable states in the process, $A$ is the set of available actions, $R$ is the mapping of transitions to the immediate reward, and $P$ represents the transition probabilities (i.e., system dynamics). At any given time-step $t$, the MDP is at a state $s_t\in S$, which can represent the current configuration of simulated CAS. The reinforcement learning agent's choice of action at time $t$, $a_t \in A$ causes a transition from $s_t$ to a state $s_{t+1}$ according to the transition probability $P_{s_t , s_{t+a}}^{a_t}$. The agent receives a reward $r_t = R(s_t, a_t) \in \mathbb{R}$ for choosing the action $a_t$ at state $s_t$.

Interactions of the agent with MDP are captured in a policy $\pi$. When such interactions are deterministic, the policy $\pi: S\rightarrow A$ is a mapping between the states and their corresponding actions. A stochastic policy $\pi(s,a)$ represents the probability of optimality for action $a$ at state $s$.

The objective of reinforcement learning is to find the optimal policy $\pi^\ast$ that maximizes the cumulative reward at any time $t$, denoted by the return function $\hat{R} = \sum_{T}^{t' = t} \psi^{t'-t} r_{t'}$, where $\psi < 1$ is the discount factor that accounts for the diminishing worth of rewards obtained further in time, hence ensuring that $\hat{R}$ is bounded.

An approach to this problem is the \emph{Action-Value Function} optimization algorithm or Q-Learning. In every iteration of this technique, the optimal value of each action is calculated as the expected sum of future rewards, assuming that every action taken after the current choice follows the optimal policy. Under a given policy $\pi$, the value of an action $a$ in a state $s$ is given by the value function $Q$ defined as:

\begin{eqnarray} \label{bellman}
Q(s_t,a_t) = R(s_t,a_t)+ \psi \max_{a_{t+1}} (Q(s_{t+1},a_{t+1}))
\end{eqnarray}

The optimal $Q$ value is hence defined as: $Q^\ast (s_t, a_t) = \max_\pi Q^\pi (s_t, a_t)$, and the optimal policy is given by $\pi^\ast(s_t) = \arg\max_{a_t} Q(s_t,a_t)$.

The Q-learning method estimates the optimal action policies by using the Bellman equation $Q_{i+1} = \mathbf{E}[R + \psi \max_{a_t} Q_i]$ as the iterative update of a value iteration technique. Practical implementation of Q-learning is generally based on function approximation of the parametrized Q-function $Q(s_t,a_t; \theta) \approx Q^\ast (s_t,a_t)$. A common technique for approximating the parametrized non-linear Q-function is to train a neural network whose weights correspond to $\theta$. This neural network is trained such that at every iteration $i$, it minimizes the loss function:
\begin{eqnarray}
L_i(\theta_i) = \mathbf{E}_{s_t, a_t\sim \rho(.)} [(y_i - Q(s_t,a_t,;\theta_i))^2]
\end{eqnarray}
where $y_i = \mathbf{E}[R + \psi \max_{a_{t+1}}Q(s_{t+1},a_{t+1};\theta_{i-1}) | s_t,a_t]$, and $\rho(s_t,a_t)$ is a probability distribution over states $s_t$ and actions $a_t$.

This optimization problem is typically solved using computationally efficient techniques such as Stochastic Gradient Decent (SGD). This approach allows for the problem of estimating $Q$ functions to be performed by neural network function approximators optimized via stochastic gradient descent, updating the current value $Q(s_t, a_t; \theta_t)$ towards a target value $Y^Q_{t}$. 

Once the optimal policy is obtained from the secondary simulation, it is implemented on the primary simulation to observe the impact for a user-defined number of timesteps. At this point, the new observations are fed back to the estimation algorithm to improve adversary's model of target dynamics, and derive the optimal attack policy for the updated model. This iterative process is executed until the user-defined criteria for attack success or termination are reached. At every iteration of Q-Learning, the process selects its estimation of the best possible action, which is one of the designated adversarial actions designated in the configuration of attack simulation. 

This process is formalized in Algorithm \ref{algo:exploit}. Before execution, this algorithm must be integrated with a dynamical simulation or physical prototype of the target system. Also, the user shall define a technique for estimation of dynamics, designate an attack objective, the set of permissible adversarial actions, the cost function of attack, and the criteria for termination of Q-learning. Upon execution, the algorithm iteratively observes the state of the target system, and updates its estimate of target's dynamics according to a pre-defined technique (line 5). This estimate is then used to create a simulation of target from an adversary's perspective, which is then explored via Q-learning to obtain an optimal attack policy based on current estimate (line 6). This policy is then applied to the original simulation or prototype of the target (line 7), and the simulated adversary's observation of target's state is updated according to the resulting state of the target (line 8). This process is repeated until the adversarial reward reaches the designated attack objective (line 4). 

It is noteworthy that this framework can only succeed if the attack objective is reachable from the initial state of the target, and with the defined set of actions. Otherwise, this algorithm will provide a best-effort performance in coming as close as possible to the objective. Also, the accuracy and convergence of this algorithm is heavily dependent on the dynamic estimation mechanism. The choice of estimation technique and its updating criteria must be such that the estimation errors do not consistently accumulate, and remain bounded over a large number of iterations. 

Furthermore, Algorithm \ref{algo:exploit} does not intrinsically account for constraints on execution time, therefore such limitations must be implemented within attach the cost function. Similar to the reachability criteria of optimality, if time constraints of the problem fall below the time required for reaching the optimal answer, this algorithm still performs a best-effort search for optimal attacks and potential impact. Such best-effort results are indeed representative of practical worst-case impact levels under the conditions modeled by user-defined parameters.

\begin{algorithm}[t]
	\fontsize{8pt}{8pt}
	\SetKwFunction{EstDyn}{EstimateDynamics}
	\SetKwFunction{QLearn}{QLearning}
	\SetKwFunction{Simul}{SimulateDynamics}
	\SetKwInOut{Input}{Input}\SetKwInOut{Output}{Output}
	\Input{dynamical simulation, Attack cost function $C$, objective $O$, set of actions $A$, termination criteria $X$}
	\KwData{initial target configuration $G_0$, reward/cost of attack $R$, current configuration $G$, policy $\pi$}
	\Output{optimal reward/cost of attack $R$, final configuration $G^{*}$, optimal policy $\pi^{*}(.)$}
	$R\leftarrow 0$ \\
	$G \leftarrow G_0$ \\
	Initialize $\pi$ to a random distribution\\
	\While {$R < O$}{
		$U\leftarrow$ \EstDyn{$G, X$}\\
		$R , \pi \leftarrow$ \QLearn{\Simul{$G, U, \pi$}, $G, U, X, C$}\\
		Implement $a \leftarrow \pi(G)$ \\
		Update $G$
	}
	\caption{Attack Simulation Framework} \label{algo:exploit}
\end{algorithm}

\section{Case Studies} \label{Sec:Case}
To study the performance and feasibility of our proposed framework, we investigated its application to 3 real-world CAS scenarios, namely: Inducing cascade failures in power distribution networks, destabilization of terrorist organizations, and policy manipulation in Deep Q-Learning. For each case study, we describe the objective and classify the type of attack according to the schemes introduced in Section \ref{Sec:Surf}. We then report the approach and experimental setup, and present the results in terms of quantitative impact and vulnerability.

\subsection{Cascade Failures in Power Grids}
Power distribution networks constitute a well-known instance of CAS \cite{pagani2013power} that are susceptible to cascading failures triggered by malfunctions in one or more local components, such as relays and transmission lines. In such cases, the load of a failed component is balanced onto neighboring nodes, causing them to overload and fail as well \cite{yan2017q}. In this case study, the attack objective is to analyze the maximum possible disconnection of a power network by induction of cascading failures through sequential removal of transmission lines in a simulated power grid. The case of sequential attacks on power grids is recently studied by Yan et al. \cite{yan2017q}, who also use a an approach based reinforcement learning to analyze the impact of such attacks. One major difference between the methodology of \cite{yan2017q} and this case study is the assumption of a blackbox attack in our approach, which circumvents the issues caused by modeling challenges in the study of cascading power grid failures \cite{guo2017critical}. Moreover, this case study demonstrates an instance of applying a dynamical system model to analysis of vulnerabilities in CAS.

\subsubsection{Objective and Classification}
The objective of this attack is to disconnect the minimum number of transmission lines one at a time, such that the system collapses. This attack targets the network structure to compromise the Availability dimension of CIA by implementing an adversarial action to manipulate the state of this CAS.

\subsubsection{Experiment Setup}
The benchmark network used in this experiment is a mid-size IEEE RTS-79 architecture \cite{grigg1999ieee}. This system is comprised of 24 buses, 38 transmission lines, 17 load buses, and 10 generating units, with a total generation capacity of 3405MW, and a peak load of 2850MWs. A line is considered to be alive if it operates with a load that is smaller than its capacity. Once this threshold is reached, the line fails and all of its load is distributed equally among the nodes that are directly connected to it.

The dynamical simulation was implemented in Python using the PyPSA toolbox \cite{brown2017pypsa}. Following the setup of \cite{yan2017q}, the attack objective was set to cause at least 8 lines failures, while minimizing direct disconnection of lines by the attacker, and maximizing the disconnections resulting from cascading failures. We constrained the maximum number of iterations of each simulation to 500, and repeated each full simulation 100 times. As for the estimation method, we adopted the architecture proposed in \cite{wang2017new} for a convex-based Long-Short Term Memory (LSTM) neural network to approximate the nonlinear dynamics of the power grid.

\subsubsection{Results}
\begin{figure}[h]
	\includegraphics[width=\linewidth]{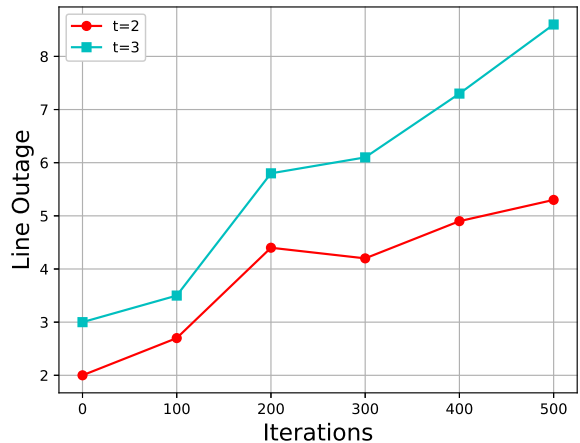}
	\caption{Induction of cascade failure in power grids via direct targeting of t=2 and t=3 lines}
	\label{FigGrid}
\end{figure}

Figure \ref{FigGrid} depicts the obtained results, avereged over 100 repetitions. It can be seen that our framework achieves an outage of 8.6 lines with only 3 direct node removals, thereby demonstrating the applicability of our framework in simulating emergent attacks in real-world CAS. Accordingly, the vulnerability measure of this network structure to node removal attacks is $\frac{1}{3} = 0.34$. 

\subsection{Destabilization of Terrorist Networks}
This case study reports our previous work in \cite{CT1}. In this work, we investigated the performance of our framework in deriving optimal destabilization policies against terrorist organizations. In the context of this study, destabilization is defined as minimizing the desire of terrorist agents to remain affiliated with the organization. Similar to the previous experiment, the choice of adversarial action in this scenario is also sequential removal of nodes (i.e., human actors). This attack aims to eliminate the spiritual and operational incentives of remaining in the organization through removal of those nodes who are vital in preserving these two aspects.

\subsubsection{Classification}
Although this is another network node removal action, but the attack surface in this scenario is the cooperation protocols of the targeted CAS. This attack aims to exploit the self-interested nature of agents by diminishing the incentive of cooperation, such that righteous breaking off from this cooperation becomes inevitable. Consequently, this attack is targeting the Integrity and Availability dimensions of CIA through active attacks on both the state and the dynamics of this CAS.

\subsubsection{Experimental Setup}
We modeled the dynamics of this CAS as a network formation game, in which the payoff function for each agent is defined as follows:

\begin{eqnarray}
U_i = \sum_{\substack{j\in N}\ 			\ j \neq i\ } G_{ij}(V_{ij}(G_{-i}, X; \theta_0)+\epsilon_{ij})\label{payoff}
\end{eqnarray}
where $G_{-i}$ is the adjacency matrix $G$ with the $i$th row deleted, and payoffs are known up to $\theta_0$. $X = (X_{ij}; i, j \in N)$ is the set of homophily vectors between all pairs $i\neq j$ obtained from profile vectors $F_i$ and $F_j$, $V_{ij}$ is the deterministic component of the payoff, and the parameter $\epsilon_{ij}\in \epsilon_i$ is the idiosyncratic shock, representing the effect of $i$'s unknown parameters on its desire to establish a link with $j$. Instances of such parameters are personal taste and psychology, and may extend to include the effects of homophily and topological parameters that are not accurately observed by the counter-terrorism entity. This factor is only known to $i$, and other players are not aware of its value. 

Consequently, the problem of estimation is simplified into estimation of payoff's parameters for each agent. The set of available data for this estimation includes automatically extracted profiles and incomplete snapshots of the network mined from open-source structured and unstructured sources. We applied a recently proposed 2-step estimation technique \cite{leung2015two} that exploits the hierarchical symmetries in the CAS to eliminate the need for detailed time-series observations of the target. 

With this estimation technique at hand, we applied the simulation framework to our extracted dataset of Al Qaeda's leadership network, with the objective of maximizing the network fragmentation $F$, defined as the proportion of mutually reachable nodes as each node is removed or unconnected from the network. Formally,

\begin{eqnarray}
F = 1 - \frac{\sum_k s_k (s_k -1)}{n(n-1)}
\end{eqnarray}

where $s$ is the size of component $k$ (i.e., groups of nodes remaining connected after removal of a node) and $n$ is the total number of nodes in the network. Values close to 1 indicate high fragmentation and values close to 0 indicate low fragmentation. As such, fragmentation is an inverse measure of the amount of connectedness or connection redundancy in a network.  

\subsubsection{Results}
\begin{figure}
	\centering
	\includegraphics[width=\linewidth]{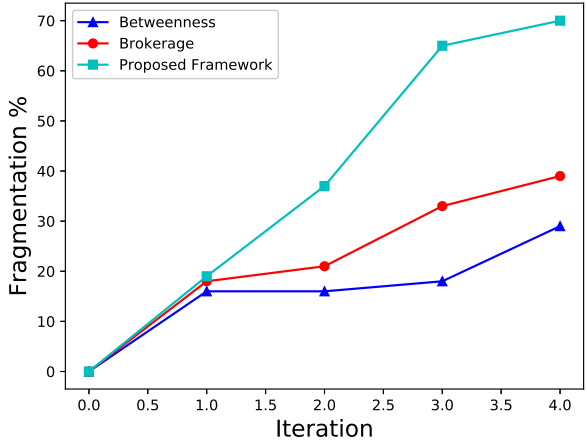}
	\caption{Performance comparison of the derived action policy}
	\label{fig:res1}
\end{figure}

Figure \ref{fig:res1} illustrates the results of implementing the action policy generated by our framework for 5 iterations, in comparison with two well-established targeting techniques in network targeting: elimination of the node with highest betweenness centrality, and elimination of those with highest brokerage values at each iteration. It is shown that the proposed technique achieves a much higher fragmentation in all steps of the process, culminating in a 71\% fragmentation after 4 node removals. The resulting order of targeting in this experiment is as follows: 
\begin{enumerate}[{(i)}]
	\item \emph{Khalid Sheikh Mohammad} : Senior figure in propaganda operations and strategic planning of attacks, including that of September 11.
	\item \emph{Ayman al-Zawahiri}: Deputy leader of Al Qaeda at the time, who has now replaced Bin Laden as the leader.
	\item \emph{Osama Bin Laden}
	\item \emph{Abu Musab al-Zarqawi} : Faction commander and senior military figure at the time, who later became the leader of Al Qaeda in Iraq.
\end{enumerate}
In the attack based on betweenness, the targeting sequence is Bin Laden first, followed by Zawahiri, Mohammad Ata (operation leader for September 11 attacks), and Abu Gatada. Also, the targeting sequence of brokerage-based attack is ordered as: Zawahiri, Abu Qatada, Bin Laden, and Ibrahim Maidin (military leader of Jemaah Islamiah in Singapore).

\begin{figure}
	\centering
	\includegraphics[width=\linewidth]{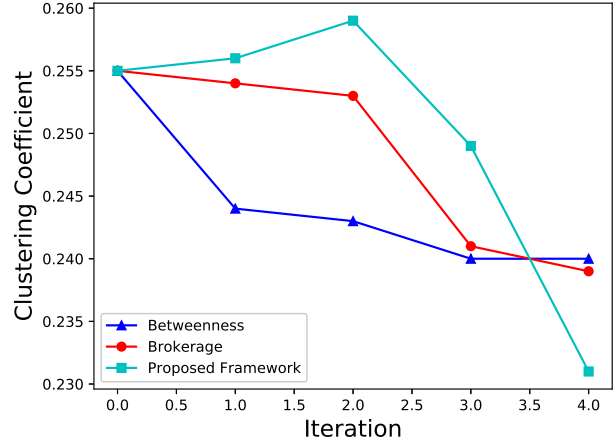}
	\caption{Variation of global clustering coefficient}
	\label{fig:res2}
\end{figure}

Due to the unavailability of ground truths in the public domain, this experiment is restricted to observations and interpretive evaluation. One interesting observation is that this policy does not recommend the targeting of Bin Laden as the first action, which as is the case today, would only lead to his replacement by Zawahiri without any major impact on the individual utilities of lower members. This policy begins by removing those nodes whose replacement leads to significant drops in network's performance, which in turn reduces the benefits of remaining in or joining the network for other members. Consequently, targeting the top leader leads to less effective replacements and network configurations, which may either dissolve on its own, or can be targeted with greater ease than the original network. This weakening of ties can be observed in the sparsity and diminishing clustering, quantified via changes in the global clustering coefficient, as depicted in Figure \ref{fig:res2}.

Accordingly, the vulnerability of Al Qaeda's network structure to node removal attacks is $\frac{1}{4} = 0.25$.

\subsection{Policy Induction in Deep Reinforcement Learning}
The emerging paradigm of deep Reinforcement Learning (RL) \cite{li2017deep} demonstrates the defining characteristics for CAS: training of deep RL is governed by the nonlinear dynamics of neural networks and interactions with its environment, the behavior of deep RL is an emergent result of local interactions between the hierarchical layers of deep networks, the policy and actions of deep RL adapt in response to changes in the environment, and the deep neural networks of this system self-organize through adjustment of inter-layer weights. Consequently, deep RL can also be subject to dynamical attacks.

To demonstrate the vulnerability of deep RL to such attacks, in \cite{behzadan2017vulnerability} we present a DDDAS-based model of vulnerabilities in such systems, and report the performance of our framework against Deep Q-Networks (DQNs) through manipulation and induction of adversarial policies in these systems at training time. In this attack, we utilize adversarial examples \cite{papernot2016practical} to manipulate the environmental feedback of DQN, and lead it towards learning our adversarial policy instead of one that satisfies the original objectives of the DQN.

The procedure of this attack can be divided into the two phases of initialization and exploitation. The initialization phase implements processes that must be performed before the target begins interacting with the environment, which are:

\begin{enumerate}
	\item Train a DQN based on attacker's reward function $r'$ to obtain the adversarial policy $\pi^\ast_{adv}$
	\item Create a replica of the target's DQN and initialize with random parameters 
\end{enumerate}

\begin{figure}[!h]
	
	\centering
	
	\includegraphics[width=\linewidth]{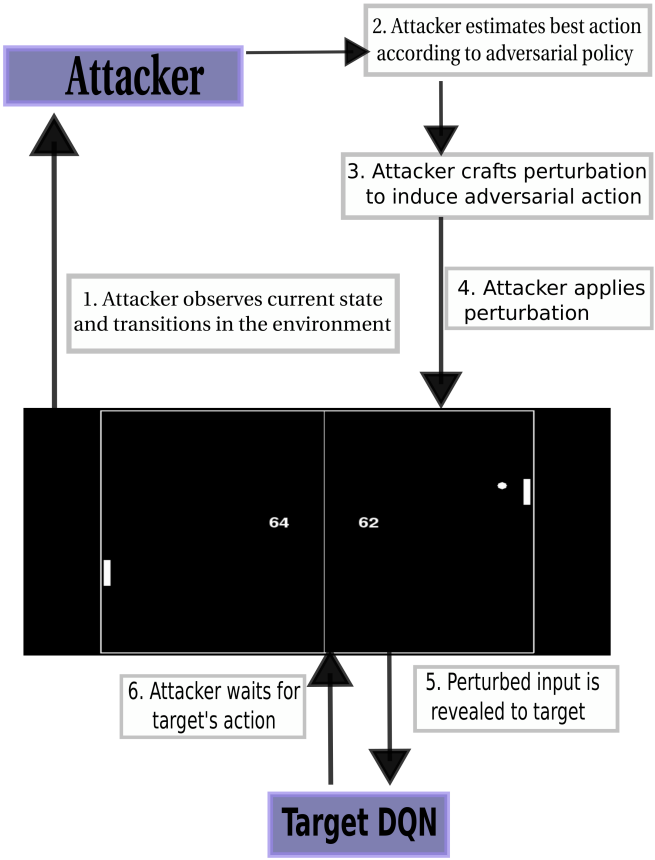}
	
	\caption{Exploitation cycle of policy induction attack}
	
	\label{fig:exploit}
\end{figure}

The exploitation phase implements the attack processes such as crafting adversarial inputs. This phase constitutes an attack cycle depicted in Figure \ref{fig:exploit}. The cycle initiates with the attacker's first observation of the environment, and runs in tandem with the target's operation. 

\subsubsection{Classification}
This attack targets the actuation functions of the CAS by compromising the Integrity and Availability dimensions of CIA through exploitation of adversarial examples to manipulate both the state and dynamics of the target.

\subsubsection{Experimental Setup}
We examine the targeting of Mnih et al.'s DQN designed to learn Atari 2600 games \cite{mnih2015human}. In our setup, we train the network on a game of Pong. The game is played against an opponent with a modest level of heuristic artificial intelligence, and is customized to handle the delays in DQN's reaction due to the training process. The game's back-end provides the DQN agent with the game screen sampled at 8Hz, as well as the game score (+1 for win, -1 for lose, 0 for ongoing game) throughout each episode of the game. The set of available actions $A = \{UP, DOWN, Stand\}$ enables the DQN agent to control the movements of its paddle.

Similar to the original architecture of Mnih et al. \cite{mnih2015human}, this input is first passed through two convolutional layers to extract a compressed feature space for the following two feed-forward layers for Q function estimation. The discount factor $\gamma$ is set to $0.99$, and the initial probability of taking a random action is set to $1$, which is annealed after every $500000$ actions. The agent is also set to train its DQN after every $50000$ observations.

In this experiment, we consider an adversary whose reward value is the exact opposite of the game score, meaning that it aims to devise a policy that maximizes the number of lost games. To obtain this policy, we trained an adversarial DQN on the game, whose reward value was the negative of the value obtained from target DQN's reward function. With the adversarial policy at hand, a target DQN was setup to train on the game environment to maximize the original reward function. The game environment was modified to allow perturbation of pixel values in game frames by the adversary. A second DQN was also setup to train on the target's observations to provide an estimation of the target DQN to enable blackbox crafting of adversarial example. At every observation, the adversarial policy obtained in the initialization phase was consulted to calculate the action that would satisfy the adversary's goal. Then, the JSMA algorithm was utilized to generate the adversarial example that would cause the output of the replica DQN network to be the action selected by the adversarial policy. This example was then passed to the target DQN as its observation.

\subsubsection{Results}
\begin{figure}[!h]
	\vspace{-8mm}
	\centering
	
	\includegraphics[width=\linewidth]{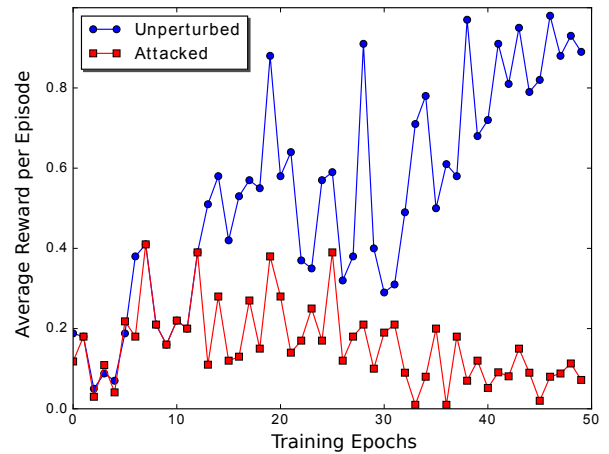}
	\caption{Comparison of rewards between unperturbed and attacked DQNs}
	
	\label{fig:plt3}
\end{figure}
Figure \ref{fig:plt3} compares the performance of unperturbed and attacked DQNs in terms of their average reward values per episode. It can be seen that the reward value for the targeted DQN agent rapidly falls below the unperturbed case and maintains the trend of losing the game throughout the experiment.

The vulnerability of DQN to policy induction attacks can be expressed as the inverse of number of epochs before divergence of average reward from the unperturbed trajectory, which is $\frac{1}{12} = 0.083$.  

\section{Discussion on Mitigation Techniques} \label{Sec:Defense}
The complexity and scope of CAS gives rise to an everlasting potential for discovery of novel and unprecedented vulnerabilities. As a result, comprehensive analysis of their resilience to adversarial actions cannot solely rely on evaluation of pre-defined lists of attack types and vectors. Consequently, such analyses must determine the underlying parametric relations and bounds which lead to CAS designs that are guaranteed to satisfy the desired criteria for reliability and security. Also, this level of resilience needs to be balanced against cost and operational specifications. Therefore, the problem of choosing optimal resilience criteria and parametric bounds translates into an iterative optimization problem. A further challenge in this analysis is to determine the temporal depth of tracing the impact of parametric changes, i.e., how far into the future is to be analyzed in order to verify the safety of tested changes. A prominent instance of this challenge is the domain of AI safety, which is concerned with the effects of long-term learning and cumulative autonomy on safe and secure operation of intelligent agents.

With regards to attack detection, the distributed nature of CAS gives rise to a major challenge in monitoring the state of the system and detection of attacks. Feasible detection mechanisms must provide the means for dissemination of local observations into a network that may be jammed, compromised, or not homogeneously trustworthy. Therefore, information sharing and incorporation of received data into attack detection mechanisms have to follow strategic and selective procedures. Also, dissemination of state information must follow protocols that minimize communication overheads, while providing reliable transmissions in networks under attack. Adoption of similar developments in such fields as cognitive radios \cite{sharma2015advances}, wireless sensor networks \cite{butun2014survey}, and Internet of Things \cite{gendreau2016survey} may prove useful in explorations of this area. 


Building on this step, a further venue of pursuit is the formal and numerical investigation of the impact of constituent element and system parameters on the resilience of CAS. Through parametric analysis of homeostasis conditions in generic models of CAS, this direction of work enables the establishment of absolute and relative parametric bounds within which a CAS remains resilient. One of the potential pursuits of this venue is to establish bounds on the initial conditions required for the emergence of resilient CAS, such as the number of constituent elements, required redundancies, and other parametric rules for schemata that give rise to the emergence of resilience. This analysis will enable a further formal investigation into the balance of resistance and adaptivity of systems with their feasibility in terms of efficacy, performance, real-time responsiveness, and energy efficiency. Achieving these objectives in large-scale CAS will require extending the models of dynamics established in Section \ref{Sec:Model} into tractable models that are better-suited for analysis of high-dimensional nonlinear dynamics. Promising venues of investigation include modern variants of MDP modeling and reinforcement learning, path integrals,  genetic algorithms, mean-field game theory, and operad theory, to analyze reachability, controllability, convergence, and phase transitions in high-dimensional state trajectories of CAS. 

Inspired by the biological phenomena of threat detection and alerting of cohorts in biological systems and societies, a further venue is to investigate the equipment of elements in CAS with intrinsic mechanisms for self-regulation and identification of ongoing attacks, thereby greatly enhancing the resilience and elasticity of such systems against adversarial manipulations. This thrust may investigate the mechanisms of anomaly detection and coalition formation that enable cooperative detection of attacks through distributed information sharing and processing. A highly useful result of this work can be the development of self-organizing mechanisms and schema that produce such functionalities as emergent behaviors of the system. A major inspiration for this study is the human immune system, which itself is a CAS whose emergent behavior is to detect, announce, and defend against attacks. Cells in the immune system perform distributed anomaly detection based on simple learning and memory retainment mechanisms, and modulate their individual and coordinated response according to the continual updates of the memory by the learning mechanism. This calls for a comprehensive study into the feasibility of such mechanisms for development of emergent defense mechanisms in CAS. This study may also investigate the employment and enhancement of multi-agent reinforcement learning and transfer learning techniques as mechanisms for adaptive learning of optimal actions in the presence of persistent and dynamic anomalies. A further direction of this task is to analyze the feasibility of embedding dedicated attack detection and mitigation nodes, and to establish design rules for balanced and optimal size, distribution, and signaling of such nodes in resilient CAS. 

\section{Conclusion} \label{Sec:Conclusion}
We introduced the paradigm of adversarial attacks targeting the nature of dynamics in Complex Adaptive Systems (CAS). Aiming to develop a comprehensive foundation for analysis of such attacks, we proposed three approaches to the modeling of CAS as dynamical, data-driven, and game-theoretic systems. We developed suitable definitions of attack, vulnerability, and resilience in the context of CAS Security, and introduced three schemes for classifying threats based on security dimensions, data-driven abstraction, and fundamental functionalities of CAS. Building on this foundation, we proposed a framework for simulation and analysis of attacks on CAS, and demonstrated its performance in vulnerability analysis of power grids, terrorist networks, and deep reinforcement learners. These case studies also demonstrate the need for novel techniques and methodologies for threat detection and mitigation in both natural and engineered CAS. To facilitate the search for such techniques, we also presented a discussion on promising approavenues for future research in analysis and design of resilient complex adaptive systems.   

\section*{Acknowledgements}
This work was supported by the National Science Foundation (NSF) (NSF-CRII-CPS-1743490). Any opinions, findings, and conclusions or recommendations expressed in this material are those of the author and do not necessarily reflect the views of the NSF. 

\vspace{-10pt}
\bibliographystyle{ieeetr}
\bibliography{ref_vahid}
\end{document}